\documentclass[12pt]{article}
\usepackage{amsmath,amssymb,amsthm,color}
\usepackage{graphicx}
\usepackage{cite}
\usepackage{caption}
\usepackage{epsfig}
\usepackage{epstopdf} 
\renewcommand{\baselinestretch}{2}
\begin{document}
%
\title{Diversified essential properties in halogenated graphenes}
\author{
\small Ngoc Thanh Thuy Tran$^{a}$, Duy Khanh Nguyen$^{a}$, Glukhova O. E.$^{b}$, Ming-Fa Lin$^{a,*}$\\
\small $^a$Department of Physics, National Cheng Kung University, Tainan 701, Taiwan \\
\small $^b$Department of Physics, Saratov State University, Saratov 410012, Russia}
\renewcommand{\baselinestretch}{1}
\maketitle

\renewcommand{\baselinestretch}{1.4}
\begin{abstract}

The significant halogenation effects on the essential properties of graphene are investigated by the first-principles method. The geometric structures, electronic properties, and magnetic configurations are greatly diversified under the various halogen adsorptions. Fluorination, with the strong multi-orbital chemical bondings, can create the buckled graphene structure, while the other halogenations do not change 
the planar $\sigma$ bonding in the presence of single-orbital hybridization. Electronic structures consist of the carbon-, adatom- and (carbon, adatom)-dominated energy bands. All halogenated graphenes belong to hole-doped metals except that fluorinated systems are middle-gap semiconductors at sufficiently high concentration. Moreover, the metallic ferromagnetism is revealed in certain adatom distributions. The unusual hybridization-induced features are clearly evidenced in many van Hove singularities of density of states. The structure- and adatom-enriched essential properties are compared with the measured results, and potential applications are also discussed.

\vskip 1.0 truecm
\par\noindent

\vskip1.0 truecm

\par\noindent  * Corresponding authors. {~ Tel:~ +886-6-2757575-65272 (M.F. Lin)}\\~{{\it E-mail address}: mflin@mail.ncku.edu.tw (M.F. Lin)}
\end{abstract}
\pagebreak
\renewcommand{\baselinestretch}{2}
\newpage

\section{Introduction}
\bigskip

Chemical modifications can dramatically change the essential properties of graphene systems. The adatom-doped graphenes have attracted a lot of theoretical\cite{praveen2015adsorption,yang2010two,nakada2011migration} and experimental \cite{wu2009organic,wang2013situ,wang2010nitrogen,veerapandian2012synthesis} researches. The electronic properties, one of important topics in physics, chemistry, and materials, are greatly diversified by various adatom adsorptions. This will lead to high potentials in near-future applications as a result of the tunable and remarkable electronic properties, e.g., optoelectronics,\cite{loh2010graphene,wu2009organic} energy storage,\cite{porro2015memristive,wang2013situ} and sensors.\cite{wang2010nitrogen,veerapandian2012synthesis} Halogenated graphenes are good candidates for studying the diverse phenomena. Halogen atoms possess very strong electron affinities, so that the significant chemical bondings with carbon atoms will play a critical role in geometric structures and electronic properties. Whether five kinds of halogen adatoms (X = F, Cl, Br, I, and At) exhibit the similar features or the important differences deserves a systematic investigation.

The fluorinated graphenes are commonly synthesized using a two-step process, fluorination to obtain fluorinated graphite and then exfoliation to achieve monolayer one. There are several methods for fluorination, such as direct gas,\cite{cheng2010reversible,mazanek2015tuning} plasma,\cite{wang2014fluorination,sherpa2014local} photochemistry\cite{gong2013photochemical,lee2012selective} and hydrothermal reaction.\cite{samanta2013highly,wang2012synthesis} The second step could be done by liquid-phase exfoliation,\cite{zhang2013two} modified Hummer's exfoliation\cite{romero2013fluorinated}, thermal exfoliation\cite{dubois2014thermal}, and solvothermal exfoliation.\cite{sun2014solvothermally} Monolayer F-doped graphene could also be directly synthesized from the chemical reaction method by heating the mixture of graphene sheet and XeF$_2$.\cite{jeon2011fluorographene,robinson2010properties} Moreover, the chlorinated\cite{li2011photochemical,wu2011controlled}, brominated\cite{jankovsky2014towards} and iodinated\cite{yao2012catalyst} graphenes have been produced by the similar methods. However, the experimental synthesis on the astatine-doped graphenes is absent up to now. This might be due to the weak At-C bondings or the small binding energies.

A lot of experimental\cite{nair2010fluorographene,jeon2011fluorographene} and theoretical\cite{medeiros2010dft,sahin2012chlorine} studies show that fluorinated graphenes belong to the unusual semiconductors or the $p$-type metals, depending on the fluorination conditions. The large energy gaps due to the high fluorination are clearly evidenced in the experimental measurements of electrical resistances,\cite{nair2010fluorographene,cheng2010reversible} optical transmissions,\cite{nair2010fluorographene,jeon2011fluorographene} and photoluminscence spectra.\cite{jeon2011fluorographene}. On the other hand, there are only few theoretical studies on the Cl-, Br-, and I-doped graphenes\cite{medeiros2010dft,sahin2012chlorine}. From few theoretical calculations,\cite{medeiros2010dft,sahin2012chlorine} the (Cl,Br,I,At)-absorbed graphenes are predicted to be in sharp contrast with the fluorinated systems, including the geometric, electronic and magnetic properties. The main differences might lie in the critical orbital hybridizations of the halogen-C bonds. Obviously, it is worthy of conducting a systematic investigation on five kinds of halogenated graphenes, in which the critical mechanisms responsible for the diversified properties and the important differences among them will be explored in detail.

The geometric, electronic and magnetic properties are studied for all the halogen-adsorbed graphenes using first-principles calculations. The dependence on the type, concentration and arrangement of halogen adatoms are investigated extensively. Binding energies, bond lengths, buckled structures, carbon- or adatom-dominated energy bands, spin-density distribution, spatial charge distribution, and density of states (DOS) are included in the calculations. Apparently, there exist the diversified essential properties, covering the opening of band gap or the distortion of the Dirac-cone structure, the metallic behaviors due to free holes, the creation of the adatom-dominated or (adatom,C)-co-dominated energy bands, the degeneracy or splitting of the spin-related energy bands, the multi- or single-orbital hybridizations in halogen-C bonds, as well as ferromagnetism and non-magnetism. They are further reflected in a lot of special structures of DOS. Such properties are strongly affected by the distinct kinds of adatoms. The theoretical calculations are compared with the experimental measurements on geometric structures and electronic properties; furthermore, the potential applications are also discussed.

\section{Computational methods}
\bigskip

In this work, the essential properties are studied by the first-principle density functional theory using the Vienna ab initio simulation package.\cite{kresse1996efficient} The exchange-correlation energy due to the electron-electron interactions is calculated from the Perdew-Burke-Ernzerhof functional under the generalized gradient approximation.\cite{perdew1996generalized} The projector-augmented wave pseudopotentials are employed to evaluate the electron-ion interactions.\cite{blochl1994projector} The wave functions are built from the plane waves with a maximum energy cutoff of $500$ eV. The spin configurations are taken into account for the adatom-adsorbed graphene systems. The vacuum distance along the z-axis is set to be $15$ {\AA} for avoiding the interaction between two neighboring cells. The first Brillouin zone is sampled in a Gamma scheme along the two-dimensional periodic direction by $12\times\,12\times\,1$ k points for structure relaxations, and by $100\times\,100\times\,1$ for further calculations on electronic properties. The convergence criterion for one full relaxation is mainly determined by setting the Hellmann-Feynman forces smaller than $0.01$ eV/{\AA} and the total energy difference of $\Delta\,E_t < 10^{-5}$ eV.

The various  atomic orbitals of carbons and adatoms are included in the first-principles calculations simultaneously, so that the dramatic changes of chemical bondings in honeycomb structure and the critical hybridization of distinct adatoms could be investigated in detail. The single- or multi-orbital chemical bondings are obtained from the detailed examinations on the atom-dominated energy bands, the spatial distributions for charge and charge difference, and the orbital-projected DOS. Such bondings will play a critical role in the essential properties so that they can account for the unusual features of halogenated graphenes. In addition, they are very useful in determining the 
various hopping integrals (the parameters) in the tight-binding model. The above-mentioned viewpoint is suitable for any condensed-matter systems. 

\section{Results and discussions}
\bigskip

Under the various halogenations, the planar/buckled structures, the metallic/semiconducting behaviors, the non-magnetism/ferromagnetism, the critical orbital hybridizations, and  the van Hove singularities are worthy of a systematic investigation. The effects due to distinct adatoms, concentrations, and distributions are explored in detail. The up-to-date experimental verifications are also discussed.

\subsection{Geometric structures}
\bigskip

Recently, theoretical and experimental researchers have been interested in investigating the geometric properties of halogenated graphenes, especially for fluorinated graphene. The geometric structures of halogen-adsorbed monolayer graphene are shown in Fig. 1. The top site is the most stable position compared to the hollow and bridge ones, consistent with the previous studies.\cite{medeiros2010dft,xu2015electronic} The optimized X-C bond lengths of halogenated graphenes grow with the increasing atomic number. For example, at the same concentration (${3.1\%}$ in Table I), they change from 1.57 {\AA} to 3.72 {\AA} during the variation of F$\rightarrow$At t. The F-C bond lengths are much smaller than the other X-C ones.  Flourination can create the buckled structures (Fig. 1(g)), in which carbon atoms in the F-C bonds deviate from the graphene sheet along the $z$-direction (heights of 0.32-0.48 {\AA} in Table 1). The $\pi$ and $\sigma$ bondings of graphene are expected to exhibit the drastic changes (discussed later). On the contrary, the other halogenated graphenes keep the planar structures even under high concentrations. In addition, the configuration parameters also depend on the adatom distribution, as shown for the 50\% cases of single- and double-side F doping (Table 1). The above-mentioned characteristics clearly indicate that the F adatom has the strongest bonds with C atoms among the halogenated graphenes. 

The binding energy energy ($E_b$) due to halogenation can account for the unusual geometric structures. $E_b$ of n adsorbed halogen (X) adatoms is expressed as $E_b$ = ($E_{tot}$ - $E_{gra}$ - n$E_X$)/n, where $E_{tot}$, $E_{gra}$, and $E_X$ are the total ground state energies of the halogenated graphene, pristine graphene, and isolated X adatom, respectively.  As illustrated in Table 1, the F adatom presents the largest binding energy, implying the most stable compared to the other halogen ones. The stability of halogenated graphene decreases in the ordering of  F $>$ Cl $>$ Br $>$ I $>$ At, which agrees well with other theoretical studies.\cite{medeiros2010dft,zbovril2010graphene} It should be noticed that the Cl-, Br-, I- and At-absorbed graphenes possess the meta-stable configurations with the adatom heights close to those of F. This configuration presents a smaller binding energy compared to the most stable one, e.g., the chlorinated systems (Table 1).\cite{medeiros2010dft,klintenberg2010theoretical} The reduced ground state energy mainly results from the weakened sp$^2$ bondings in the longer C-C bonds. The very strong orbital hybridizations in Cl-C bonds can destroy the Dirac-cone structure and create an observable energy gap (${E_g=1.41}$ eV at 100\%). But for the optimal configurations, these four kinds of halogenated systems exhibit the metallic behavior with free holes even at the saturated adsorption (discussed in Fig. 3). In short, the fluorinated graphenes sharply contrast to the other halogenated ones in geometric configurations, and so do the essential electronic properties.

It is worthy of conducting a comparison for the adatoms with strong affinities in geometric structures. There exist certain important similarities and differences among the halogenated, hydrogenated and oxidized graphenes. The former two present the optimal top-site positions,\cite{medeiros2010dft,xu2015electronic,yang2010two} whereas the last one is prefered to be adsorbed at the bridge site.\cite{saxena2011investigation} Graphene oxides possess the lowest binding energy, indicating the relatively easy synthesis in experimental laboratories. As to the adatom-carbon bond lengths, they are, respectively, longest and shortest in the non-F halogenated and hydrogenated graphenes.\cite{yang2010two,medeiros2010dft,nakada2011migration} Fluorination, hydrogenation and oxidization can create the buckled graphene structures, depending on the adatom concentration and distribution. However, the non-F halogenation remains the planar sheet. Apparently, the main features of geometric structures are closely related to the critical  chemical bondings in single- or multi-orbital hybridizations. They are further reflected in electronic structures.


\subsection{Electronic properties and magnetic configurations}
\bigskip

The two-dimensional (2D) band structures along high symmetry points are useful for examining the main features of electronic properties. Monolayer graphene has a Dirac-cone structure at the K point (Fig. 2(a)), owing to the extended $\pi$ bondings of 2p$_z$ orbitals in a hexagonal sheet. The linear energy bands are getting into the parabolic dispersions in the increase of state energy, e.g., the parabolic bands near the  $M$ point (the saddle point in the inset). Furthermore, the $\sigma$ bands, which arise from the sp$^2$ bondings, are initiated from the $\Gamma$ point at the  deeper energy. For the sufficiently low concentrations, halogen adatoms will donate free holes by the very strong affinity, leading to the distortion of the Dirac-cone structure. This relies on the strength of orbital hybridization in halogen-C bonds, as shown for halogenated graphenes in Figs. 2(b)-2(f). The Fermi level is situated at the valence Dirac cone, so that free holes exist between the valence Dirac point and the Fermi level (${E_F=0}$). The linearly intersecting bands (Fig. 2(a)) in monolayer graphene become parabolic bands with two separated Dirac points for the F-adsorbed system (Fig. 3(b)), while the other halogenated graphenes only exhibit the slightly distortions (Figs. 2(c)-2(f)). This clearly illustrates the more complicated orbital hybridizations in F-C bonds, being attributed to the lower adatom height. Such hybridizations induce the thorough changes of all energy bands. Apparently, the quasi-rigid blue shifts of the pristine energy bands are absent. Moreover, the halogen-C bond strength can determine whether the (halogen,C)-co-dominated or halogen-dominated energy bands survive. The rather strong F-C bonds cause the (F,C)-co-dominated valence bands to be located in $-$3.5 eV${\le\,E^v\le\,-2.5}$ eV. On the contrary, the Cl-, Br, I- and At-dominated bands are very similar to one another, in which they are characterized by the almost flat bands in $-$1 eV${\le\,E^v\le\,-2.5}$ eV. The weak mixing with the C-dominated $\pi$ bands directly reflects the bonding strength. The significant splitting of the spin-up and spin-down energy bands is revealed near $E_F$ (black and red curves); furthermore, the largest energy spacing reaches $\sim$ 0.5 eV. The former has more occupied electronic states (black curves), leading to the ferromagnetic spin configuration.

Electronic and magnetic properties are dramatically altered by graphene halogenations, such as the Dirac-cone structure, the Fermi level, the free carrier density, the energy gap, the adatom-dominated bands, and the spin configurations. Electronic structures of fluorinated graphenes strongly depend on the concentration and distribution of adatoms, as shown in Figs. 3(a)-3(d). The Dirac-cone structure thoroughly vanishes under the full fluorination (${100\%}$ in Fig. 3(a)). The absence of linear bands arises from the overall F-C bondings related to C-2p$_z$ orbitals; that is, the very strong F-C bonds suppress the $\pi$ bondings thoroughly. A direct energy gap of 3.12 eV appears at the $\Gamma$ point, in agreement with the previous theoretical calculations\cite{medeiros2010dft,robinson2010properties} and experimental measurements.\cite{nair2010fluorographene} It is related to the highest (F,C)-co-dominated valence bands and the lowest C-dominated conduction bands. Such energy bands are associated with the passivated carbon atoms, as indicated from orange circles. All valence bands are co-dominated by F and C, in which most of them are built from their 2p$_x$+2p$_y$ orbitals (discussed in Fig. 5(a) for DOS). Furthermore, energy dispersions (band widths) are sufficiently strong (wide) in $-$4 eV${\le\,E^v\le\,-1.56}$ eV, but become weak at deeper energies. These clearly illustrate the strong orbital interactions in F-F, C-C and F-C bonds. The former two can create energy bands, leading to the hybridized valence bands by the last one. With the decrease of fluorination, the semiconducting or metallic behaviors are mainly determined by the C-dominated energy bands (Figs. 3(b)-3(d)), since the $\pi$-electronic structure (the 2p$_z$-orbital bonding) is gradually recovered. The F-F bonding quickly declines, while the strength of F-C bond almost keeps the same, as implied from the weakly dispersive (F,C)-co-dominated bands in $-$4 eV${\le\,E^v\le\,-2.5}$ eV even at lower concentrations (Fig. 3(d)). In addition, hole doping and ferromagnetism might coexist under certain concentrations and distributions (Fig. 3(d)).

As for the other halogenated graphenes (Cl- and Br-doped graphenes), they possess the metallic energy bands under various adatom adsorptions, as shown in Figs. 3(e)-3(j). Free carriers can occupy the Dirac-cone structure, as well as the halogen-created energy bands with wide band-widths, at sufficiently high concentrations. This clearly indicates the coexistence of the $\pi$ bonding in carbon atoms and the significant atomic interactions among halogen adatoms. The latter are largely reduced in the decrease of concentration, so that halogen-dependent energy dispersions become very weak, and they are below or near the Fermi level (Figs. 3(g) and 3(j)). That is, free carriers only exist in the Dirac cone at lower concentrations. The drastic changes in the halogen-induced band widths are also revealed in the concentration-dependent DOS (Fig. 5).

The spin-density distributions could provide more information about electronic and magnetic properties. The halogenated graphenes might exhibit the ferromagnetic spin configuration, depending on which type of adatoms and adsorption positions. The fluorinated systems, as shown in Figs. 3(d), have the spin-split energy bands across the Fermi level simultaneously; that is, they present the metallic ferromagnetisms. Such bands are dominated by carbon atoms (2p$_z$ orbitals), but not F adatoms. This is clearly identified from the spin-density distributions near C atoms (Fig. 4(a)) and spin-split DOSs of C-2p$_z$ orbitals asymmetric about the Fermi level (Figs. 5(c)). The occupied/unoccupied carrier densities in the spin-up and spin-down bands differ from each other (the black and red curves in Fig. 3), in which their difference will determine the strength of the net magnetic moment (e.g., 1.94 $\mu_B$ and 0.22 $\mu_B$ for single-side 50\% and 16.7\% in Table 1, respectively). On the other hand, the metallic ferromagnetisms in other halogenated graphenes are closely related to the adatom-dominated spin-split energy bands below and near the Fermi level (Figs. 3(g)-3(f); 3(g) \& 3(j)). This is consistent with the spin-density distributions around adatoms (Figs. 4(b)-4(d)), and the asymmetric and large low-energy DOS of adatom orbitals (Figs. 5(g) \& 5(j)). The halogen-created magnetic properties on graphene surface could be examined using the spin-polarized STM.\cite{serrate2010imaging,wulfhekel2007spin}

\subsection{Chemical bondings}
\bigskip

The main characteristics of electronic structures are directly reflected in the DOS, as shown in Fig. 5. The orbital-projected DOS can clearly illustrate the orbital hybridizations in X-X, C-X and C-C bonds. There are four type of special structures in DOS, the V-shaped structures, the symmetric peaks in the logarithmic and delta-function-like forms, and the shoulders, depending on 
the critical points (the band-edge states) in the  energy-wave-vector space. For pristine graphene, the low-lying structures, as shown in the upper inset of Fig. 5(a), are due to the 2p$_z$-2p$_z$ bondings among C atoms (red curve). DOS presents a linear energy dependence near ${E=0}$ and vanishes there, illustrating the characteristic of a zero-gap semiconductor. The $\pi$ and $\pi^*$ logarithmic-form peaks, respectively, appear at ${E=-2.2}$ eV and ${2.3}$ eV.\cite{neto2009electronic} Such symmetric peaks arise from the saddle points of parabolic bands near the M point (the dashed curve in Fig. 2(a)). Moreover, the shoulder structures, which are associated with the (2p$_x$,2p$_y$) orbitals, come to exist at deeper energy (${\sim\,-3}$ eV). They correspond to the extreme band-edge states of parabolic dispersions near the $\Gamma$ point (Fig. 2(a)). The lower-energy DOS is dramatically altered after halogenation. Under the full fluorination (Fig. 5(a)), the V-shaped structure and two prominent symmetric peaks are absent, since the overall F-C interactions thoroughly destroy the $\pi$ bondings of graphene. Instead, there exist an energy gap  of 3.12 eV centered at $E_F$ and several F-dominated special structures, a result of the F-F bondings in a wide energy range below the Fermi level. (2p$_x$,2p$_y$) orbitals of fluorine and carbon atoms dominate the valence-state DOS at ${E<-1.56}$ eV (pink and blue curves). They can create the shoulder and peak structures simultaneously; furthermore, the band widths related to them are more than 3 eV. These clearly illustrate the coexistence of strong F-F, C-C and F-C bonds. In addition, the 2p$_z$ orbitals only make contributions at very deep energies (the lower inset in Fig. 5(a)). With the reduce of fluorination, there are more special structures in DOS, as indicated in Figs. 5(b)-5(d), being attributed to the narrower energy widths of F-dependent valence bands, the gradual recovery of carbon $\pi$ bondings (red curves), and the partial contributions of F-2p$_z$ orbitals (cyan curves). Specifically, F-(2p$_x$,2p$_y$) orbitals can make significant contributions even for the sufficiently low concentration; however, they are revealed at narrow energy range, e.g., certain sharp peaks. Such structures result from the weakly dispersive (F,C)-co-dominated energy bands (Fig. 2(b)). The enhancement of $\pi$-band width becomes the critical factor in determining the magnitude of energy gap and the metallic behavior. For the low concentration, a finite DOS near $E=0$ is combined with one dip structure (arrows in Fig. 5(c)) or a zero plateau (Fig. 5(d)) at the right-hand side. This is an evidence of the distorted valence Dirac cone (Figs. 3(d) \& 2(b)). It is also noticed that the metallic ferromagnetism is clearly revealed in the occupied/unoccupied spin-dependent DOSs of C-2p$_z$ orbitals near the Fermi level (the solid and dashed red curves in Fig. 5(c)), in the agreement with the carbon-dominated magnetism (Fig. 4(a)).

The other halogenated graphenes sharply contrast with fluorinated systems in certain important characteristics of DOS. They have an obvious $\sigma$ shoulder about 3 eV below $E_F$ (blue curves in Figs. 5(e)-5(j)). The $\sigma$ valence bands keep the same after halogenation, further indicating the negligible hybridizations among C-(2p$_x$,2p$_y$) orbitals and adatom ones. All systems present a dip structure near $E=0$ (arrows), in which the blue shift can reach $\sim$0.7 eV. This reflects the slightly distorted Dirac cone formed by the $\pi$ bonding of C-2p$_z$ orbitals. Furthermore, there are obvious special structures due to the $\pi$ bondings (red curves). Under the sufficiently high halogenation, the (3p$_x$,3p$_y$) and 3p$_z$ orbitals of Cl adatoms can, respectively, form the conduction and valence bands, and the valence bands (pink and cyan curves in Figs. 5(e) and 5(f)), as revealed in brominated graphenes (Figs. 5(h) and 5(i)). With the decrease of adatom concentration, their band widths decline quickly, which is illustrated by the sharp delta-function-like peaks (Figs. 5(g) and 5(j)). The (3p$_x$,3p$_y$) or (4p$_x$,4p$_y$)-dominated weakly dispersive energy bands can greatly enhance DOS near the Fermi level. Specifically, the special structures associated with the (2p$_z$,3p$_z$)/(2p$_z$,4p$_z$) orbitals appear at the same energies. Their hybridizations are the significant interactions between halogen and carbon atoms, according to the atom- and orbital-dependent features in DOSs.  Moreover, the spin-up and spin-down DOSs might be very asymmetric about the Fermi level, in which they come from the adatom p$_z$ orbitals (the inset of Figs. 5(g) and 5(j)). This is consistent with the adatom-dominated ferromagnetic metals (Figs. 4(c) and 4(d)).

In order to further comprehend the multi- or single-orbital hybridizations in halogenated graphenes, which dominate the essential properties, the spatial charge distributions have been taken into account. The charge density $\rho$ (Figs. 6(a)-6(e)) and the charge density difference $\Delta \rho$ (Figs. 6(f)-6(i)) can provide very useful information about the chemical bondings and thus explain the dramatic changes of energy bands. The latter is created by subtracting the charge density of graphene and halogen atoms from that of the composite system. $\rho$ illustrates the chemical bonding as well as the charge transfer. As F atoms are adsorbed on graphene, their orbitals have strong hybridizations with those of passivated C, as seen from the red region enclosed by the dashed black rectangles (Figs. 6(b) and 6(c)). Compared with pristine graphene (Fig. 6(a)), such strong F-O bonds lead to the deformed $\pi$ bonds (pink rectangles) and thus the serious distortion of the Dirac-cone structures (Fig. 2(a)). Between two non-passivated C atoms, $\rho$ shows a strong $\sigma$ bonding (black rectangles), being slightly reduced after the formation of F-O bonds. The deformation of $\sigma$ bonding can also be viewed in $\Delta \rho$ (black arrows). There also exist the significant F-F bonds (orange rectangles) at the sufficiently high adatom concentration. These can create the (F,C)-co-dominated energy bands, the F-induced energy bands, the more complicated $\pi$ and $\sigma$ bands, and the seriously distorted Dirac cone (Figs. 2 \& 3). The obvious spatial distribution variations on ${yz-}$ and ${xz-}$ planes clearly illustrate the multi-orbital hybridizations of (2p$_x$,2p$_y$,2p$_z$) in F-C bonds. The sp$^3$ bonding is evidenced in a buckled graphene structure, as indicated from the deformations of the $\sigma$ bonding in the nearest C atom (black arrows) and the $\pi$ bonding in the next-nearest one (pink arrows).


On the other hand, the other halogenated graphenes do not have rather strong X-C bonds and thus almost keep the same in the planar $\sigma$ bonding, as shown for Cl-doped systems in Figs. 6(d) and 6(e). Cl and C atoms are bound to each other by their $p_z$ orbitals (Figs. 6(h) and 6(i)). This critical single-orbital hybridization cannot destroy the $\pi$ bonding and the Dirac cone (Figs. 3 \& 5). This is responsible for the $p$-type doping; that is, all of them belong to metals. For high concentrations, halogen adatoms possess the (p$_x$,p$_y$)-orbital hybridizations, even creating conduction bands with many free carriers. In short, the important differences between fluorinated graphenes and other halogenated systems in the essential properties originate from the orbital hybridizations of chemical bonds. The similar orbital hybridizations are also revealed other in halogenations (not shown).

\subsection{Experimental verifications and potential applications}
\bigskip

The concentrations and distributions of halogen adatoms could be identified using experimental measurements. Both X-ray photoelectron spectroscopy (XPS) and Raman spectroscopy have confirmed  the F-concentrations of 25\% for single-side and 100\% for double-side at room temperature.\cite{robinson2010properties} Furthermore, fluorinated graphenes with double-side concentrations of 25\% and 50\% produced by exfoliating graphite fluoride in fluorinated ionic liquids are verified by transmission electron microscopy and atomic force microscopy.\cite{chang2011facile} Adjusting the F/C ratio of fluorinated graphene is very important for opening the band gap, tuning the electrical conductivity as well as optical transparency, and resolving the structural transformation. This ratio can be controlled by modifying the fluorination conditions, such as fluorination agents, temperature and time.\cite{yu2012increased,gong2012one} In additional to fluorine, XPS and energy-dispersive X-ray spectroscopy have measured the adatom concentrations in chlorinated and brominated graphenes, revealing the ranges of 18-27\% and 2-8\%, respectively.\cite{zheng2012production} Also, a 8.5\% coverage of Cl-adsorbed graphene is estimated by XPS measurement.\cite{wu2011controlled} As to the in-plane lattice constant, the expansion by fluorination is revealed in electron diffraction pattern.\cite{cheng2010reversible,nair2010fluorographene} 

Scanning tunneling microscopy (STM) is a useful tool to image the surface structure of a sample with sub-angstrom precision and atomic resolution, providing the spatially atomic distributions at the local nano-structures. STM measurements have been successfully utilized to resolve the unique geometric structures of the graphene-related systems, such as AB and ABC stackings of few-layer graphenes,\cite{vcervenka2009room,kondo2012atomic} the rippled and buckled graphene islands,\cite{bai2014creating,de2008periodically} and the adatom distributions on graphene surface.\cite{pandey2008scanning,balog2009atomic} A nanoscale periodic arrangement of O atoms is confirmed using STM.\cite{pandey2008scanning} Specifically, STM images have revealed the hydrogen adsorbate structures on graphene surface, including the top-site positions and the distinct configurations: ortho-dimers, para-dimers, and various extended dimer structures and monomers.\cite{balog2009atomic} The predicted geometric properties of halogenated graphenes, including the optimal top-site positions, the adatom-dependent heights, and the fluorination-induced buckling structure, deserve further experimental measurements. Such examinations are very useful to identify the single- or multi-orbital hybridizations in X-C bonds.

Angle-resolved photoemission spectroscopy (ARPES) is a powerful experimental technique to identify the wave-vector-dependent electronic structures. The adsorption-induced dramatic changes in band structures could be directly examined by ARPES. For example, such measurements have identified the Dirac-cone structure of graphene grown on SiC,\cite{ohta2008morphology} and observed the opening of band gap for graphene on Ir(111) through oxidation \cite{schulte2013bandgap}. The ARPES measurement on fluorinated graphenes shows a $\sim$0.79-eV redshift of the Fermi level below the Dirac point.\cite{walter2011highly} In addition to ARPES, other techniques can be frequently utilized to examine the metallic and semiconducting of halogenated graphenes are optical spectroscopies and electrical transport measurements. For chlorinated graphenes, the $p$-type doping is verified from the upshift of the graphitic G-band phonon in the Raman characterization.\cite{wu2011controlled} In addition, Br$_2$- and I$_2$-doped monolayer graphenes\cite{chu2012charge} present the $p$-type doping, since the Dirac voltage at the charge neutrality point where the 4-probe resistance is maximum is shifted to higher gate voltages in the increase of molecule concentration. A 3.8-eV energy gap of  high-concentration fluorographene is directly verified from photoluminescence spectrum and near edge X-ray absorption spectrum.\cite{jeon2011fluorographene} Also, the prominent features characteristic of the strong F-C bonds are confirmed by Fourier transform infrared spectroscopy and electron energy loss spectroscopy. A highly fluorinated graphene is verified to be transparent at visible frequencies and have the threshold absorption in the blue range, indicating a wide gap of ${E_g\ge\,3}$ eV.\cite{nair2010fluorographene} This is consistent with the high room-temperature resistance of $>$10G${\Omega}$ in the electrical measurements.\cite{nair2010fluorographene,cheng2010reversible}

The scanning tunneling spectroscopy (STS) measurements, in which the tunneling conductance (dI/dV) is proportional to the DOS, can serve as an efficient method to examine the special structures in DOS. Up to now, they have verified the diverse electronic properties in graphene nanoribbons,\cite{huang2012spatially,sode2015electronic} few-layer graphenes,\cite{yankowitz2013local,pierucci2015evidence} and adatom-adsorbed graphenes.\cite{chen2015long,gyamfi2011fe} As for the theoretical predictions, electronic properties, being diversified by the X-X, X-C and C-C bonds, could be test by ARPES on band structures and STS on DOS, such as the halogen-, carbon- and (X,C)-co-dominated energy bands, the destruction or distortion of Dirac cone, an enhanced DOS accompanied with a blue-shift dip/plateau structure near the Fermi level, a pair of gap-related shoulder structures, and halogenation-induced many special structures. The spin-polarized spectroscopies are available in examining the spin-split energy bands and the highly asymmetric DOSs near $E_F$.

The high stability and remarkable properties make fluorinated graphenes become outstanding candidates in various fields. For instance, edge-fluorinated graphene nanoplatelets could serve as high performance electrodes for lithium ion batteries and dye-sensitized solar cells.\cite{jeon2015edge} Fluorinated graphenes not only possess abundant fluorine active sites for lithium storage but also facilitate the diffusion of Li$^+$ ions during charging and discharging, leading to high-performance lithium batteries.\cite{sun2014solvothermally} Such systems could also be used as electrode materials for supercapacitors.\cite{zhao2014fluorinated} Fluorinated graphene films with graphene quantum dots are highly considered in electronic applications, mainly owing to the unipolar resistive switching effect and high current modulation.\cite{opticalcontrol2016}  Furthemore, fluorinated graphenes might have the potentials in other fields, e.g., the ink-jet printed technologies,\cite{nebogatikova2016fluorinated} the biological scaffold for promoting neuro-induction of stem cells,\cite{wang2012fluorinated} amonia detections,\cite{katkov2015backside} and biomolecules adsorptions.\cite{urbanova2016fluorinated}  Also, fluorinated graphene oxides attract a lot of attention due to its promising applications.\cite{cancer2013,vizintin2015fluorinated} They are available in biocompatible drug carrier and near infra-red laser inducible agent that can dissipate thermo-sensitive cancer cells.\cite{cancer2013} Fluorinated reduced graphene oxide is utilized as an interlayer in lithium-sulfur batteries, so that the open circuit potential, cycling stability and capacitiy are greatly improved.\cite{vizintin2015fluorinated} As for the other halogenated graphenes, there are promising applications such as Cl-graphene-based field-effect transistors,\cite{zhang2013impact} and brominatied few-layer graphenes for highly apparent conducting electrodes with low optical losses.\cite{mansour2015bromination} The further studies on functionalized halogenated graphenes are expected to achieve many interesting applications.





\section{Conclusions}
\bigskip

In summary, we use the first-principles calculations to investigate geometric structure and electronic properties of halogenated graphenes. They are very sensitive to the kind, distribution and concentration of adatom adsorptions. The rich essential properties are deduced to come from the single- or multi-orbital chemical bondings between carbons and adatoms. By the detailed analyses on the calculated results, the critical orbital hybridizations could be obtained from the atom-dominated band structures, bands, the spatial charge distributions, and the orbital-projected DOS. This theoretical framework is very useful in the future studies on the chemical doped emergent materials. Up to now, part of theoretical predictions are consistent with the experimental measurements. Most of them are worthy of closer examinations.

Halogenated graphenes exhibit the diverse and unique chemical bondings, especially for the great differences between fluorination and other halogenations. There are X-X, X-C and C-C bonds, in which the former is presented at higher concentrations. The (2p$_x$,2p$_y$,2p$_z$) orbitals of F and C have very strong hybridizations among one another. This leads to the buckled structure with the drastic change in $\pi$ bonding, the destruction or serious distortion of Dirac cone, and the absence of energy bands due to F-2p$_z$ orbitals. The (2p$_x$,2p$_y$) orbitals of F adatoms can built energy bands, combined with those of C atoms by the F-C bonds. Under the low fluorination, they become narrow (F,C)-co-dominated bands at middle energy. Moreover, fluorinated graphenes are hole-doped metals or semiconductors, depending on F-concentrations and -distributions. On the other hand, only the significant interactions of $p_z$ orbitals exist in other X-C bonds, and the $\sigma$ bonding of graphene is unchanged after halogenations. Electronic structures mainly consist of the carbon- and adatom-dependent energy bands; the mixings between the $p_z$-dominated ones are observable. The $\sigma$ bands almost keep the same, and the Dirac cone is slightly distorted. The other halogenated graphenes belong to metals. The (p$_x$,p$_y$) and p$_z$ orbitals of halogen adatoms, respectively, create the distinct energy bands. They have the weakly energy dispersions at low concentrations, and make much contribution to low-energy DOS. The unusual hybridization-induced features result in many special structures of DOS. The carbon- and adatom-dominated magnetisms are, respectively, revealed in the fluorinated and other halogenated systems. Halogenated graphenes are expected to have highly potential applications in many fields, such as lithium-related batteries,\cite{sun2014solvothermally,jeon2015edge,vizintin2015fluorinated} supercapacitors,\cite{zhao2014fluorinated} ink-jet printed technologies,\cite{nebogatikova2016fluorinated} and field-effect transistors.\cite{zhang2013impact} 

\par\noindent {\bf Acknowledgments}

This work was supported by the National Center for Theoretical Sciences (South) and the Nation Science Council of Taiwan, under the grant No. NSC 102-2112-M-006-007-MY3. 

\newpage
\renewcommand{\baselinestretch}{0.2}

\newpage
\begin{table}[htb]
\small
\caption{The calculated C-C and C-X bond lengths, heights (shift on z-axis) of passivated carbons, binding energies, energy gaps, and total magnetic moments per unit cell of halogen-absorbed graphene systems. The double-side structures are labeled with *. $\triangle$ represents the very strong Cl-C bonding.}
\label{t6}
\centering
\begin{tabular*}{1\textwidth}{@{\extracolsep{\fill}}llllllllll}
\hline
Adatom& X:C&\% & &Bond length& &Height &$E_b$ &$E_g$&M$_{tot}$\\
 & & &C-X (\AA)&1st C-C (\AA)&2nd C-C (\AA)& (\AA)&(eV)& (eV)&($\mu_B$)\\
\hline
F& 8:8$^*$ & 100 &1.38 &1.58 & &0.48& -3.00 &3.12  & 0 \\	
 & 4:8$^*$ & 50 &1.48 &1.51 &  & 0.38  & -2.71  &  2.77 &0\\
 & 4:8 & 50 &1.48 &1.5 &  & 0.28 & -1.16  & 0  &1.94\\
 & 2:8 & 25 &1.54&  1.48&1.43  &0.34 & -2.94  & 2.94 &0\\	
 & 1:6 & 16.7 & 1.53 & 1.48& 1.42&  0.32& -2.45&  0&0.22\\	
 & 1:8 & 12.5 & 1.53 & 1.49 & 1.42 &0.32  &-2.64  & 0 &0.5\\
 & 1:32 & 3.1 &1.57  &1.45  &1.42  &  0.32&-2.04 &  0&0\\	
Cl & 8:8$^*$ ($\triangle$)& 100 &  1.74& 1.76 &  & 0.51 &  -0.62& 1.41 &0\\	
& 8:8$^*$ & 100 & 3.56 &  1.48&   &0  & -0.84&  0&0\\
&4:8$^*$& 50 & 3.5& 1.46&  &0 &-1.11  & 0&0\\
&2:8& 25 &3.2&  1.44&1.44  &0 &  -1.65& 0&0\\
&1:6& 16.7 & 3.2&  1.44 &  1.44&0 & -1.16 & 0&0.56\\
&1:8& 12.5 &3&   1.44&  1.43&0 &  -1.64& 0&0.57\\
& 1:32 & 3.1 & 2.96 & 1.42 & 1.42 &0  &  -0.97 &  0&0.36\\
Br& 8:8$^*$& 100 &3.56  &  1.56 &  &0  & -0.48 & 0 &0\\
&2:8& 25 & 3.59&  1.43 & 1.43 &0 & -1.91 & 0&0\\
&1:8& 12.5 & 3.56&  1.43 &1.43  &0 & -1.43 & 0& 0.48\\
& 1:32 & 3.1 & 3.23& 1.42 & 1.42 &0  &  -0.71 & 0 &0.41\\
I& 1:32 & 3.1 &  3.59& 1.42 & 1.42 &  0&  -0.48 & 0 &0.44\\	
At& 1:32 & 3.1 & 3.72 &1.42  &1.42  & 0 & -0.36 &0  &0.5\\	
\hline
\end{tabular*}
\end{table}

\newpage \centerline {\Large \textbf {FIGURE CAPTIONS}}

\vskip0.5 truecm 
\begin{itemize}
\item[Figure 1:] Geometric structures of halogen-doped graphene for various concentrations and distributions: (a) X:C = 8:8 = 100\% (double-side), (b) X:C = 4:8 = 50\% (double-side), (c) X:C = 2:8 = 25\% (single-side), (d) X:C = 1:8 = 12.5\%, (e) X:C = 1:6 = 16.7\%, (f) X:C = 1:32 = 3.1\%, and (g) side views of F-doped systems.
\bigskip

\item[Figure 2:] Band structures in a ${4\times\,4}$ supercell for (a) monolayer graphene, and the (b) F-, (c) Cl-, (d) Br-, (e) I-, and (f) At-doped graphenes with concentration X:C = 1:32 = 3.1\%. The blue circles correspond to the contributions of adatoms, in which the dominance is proportional to the radius of circle. Also shown in the inset of (a) is that for a ${1\times\,1}$ unit cell without the zone-folding effect.
\bigskip

\item[Figure 3:] Band structures of F-, Cl- and Br-doped graphenes: (a) F:C = 8:8 = 100\% (double-side), (b) F:C = 4:8 = 50\% (double-side), (c) F:C = 2:8 = 25\% (single-side), (d) F:C = 1:6 = 16.7\% (single-side), (e) Cl:C = 8:8 = 100\% (double-side), (f) Cl:C = 2:8 = 25\% (single-side), (g) Cl:C = 1:8 = 12.5\% (single-side), (h) Br:C = 8:8 = 100\% (double-side), (i) Br:C = 2:8 = 25\% (single-side), and (j) Br:C = 1:8 = 12.5\% (single-side) concentrations. The blue and orange circles correspond to the contributions of adatoms and passivated C atoms, respectively.
\bigskip

\item[Figure 4:] The spin-density distributions with top and side views under different concentrations and distributions: (a) F:C = 16.7\%, (b) Cl:C = 12.5\%, (c) Cl:C = 3.1\%, and (d) Br:C = 3.1\%. The red isosurfaces represent the charge density of spin-up configuration.
\bigskip

\item[Figure 5:] Orbital-projected DOS of F-, Cl- and Br-doped graphenes for concentrations and distributions: (a) F:C = 8:8 = 100\%, (b) F:C = 2:8 = 25\%, (c) F:C = 1:6 = 16.7\%, (d) F:C = 1:32 = 3.1\%, (e) Cl:C = 8:8 = 100\%, (f) Cl:C = 2:8 = 25\%, (g) Cl:C = 1:32 = 3.1\%, (h) Br:C = 8:8 = 100\%, (i) Br:C = 2:8 = 25\%, and (j) Br:C = 1:32 = 3.1\%.
\bigskip

\item[Figure 6:] The spatial charge densities for: (a) pristine graphene, (b) F:C = 8:8 = 100\%, (c) F:C = 2:8 = 25\%, (d) Cl:C = 8:8 = 100\%, and (e) Cl:C = 2:8 = 25\%. The corresponding charge density differences of halogenated graphenes are, respectively, shown in (f)-(i).
\end{itemize}

\newpage
\newpage
\begin{figure}[hp]
\graphicspath{{figure}}
\centering
\includegraphics[scale=1.3]{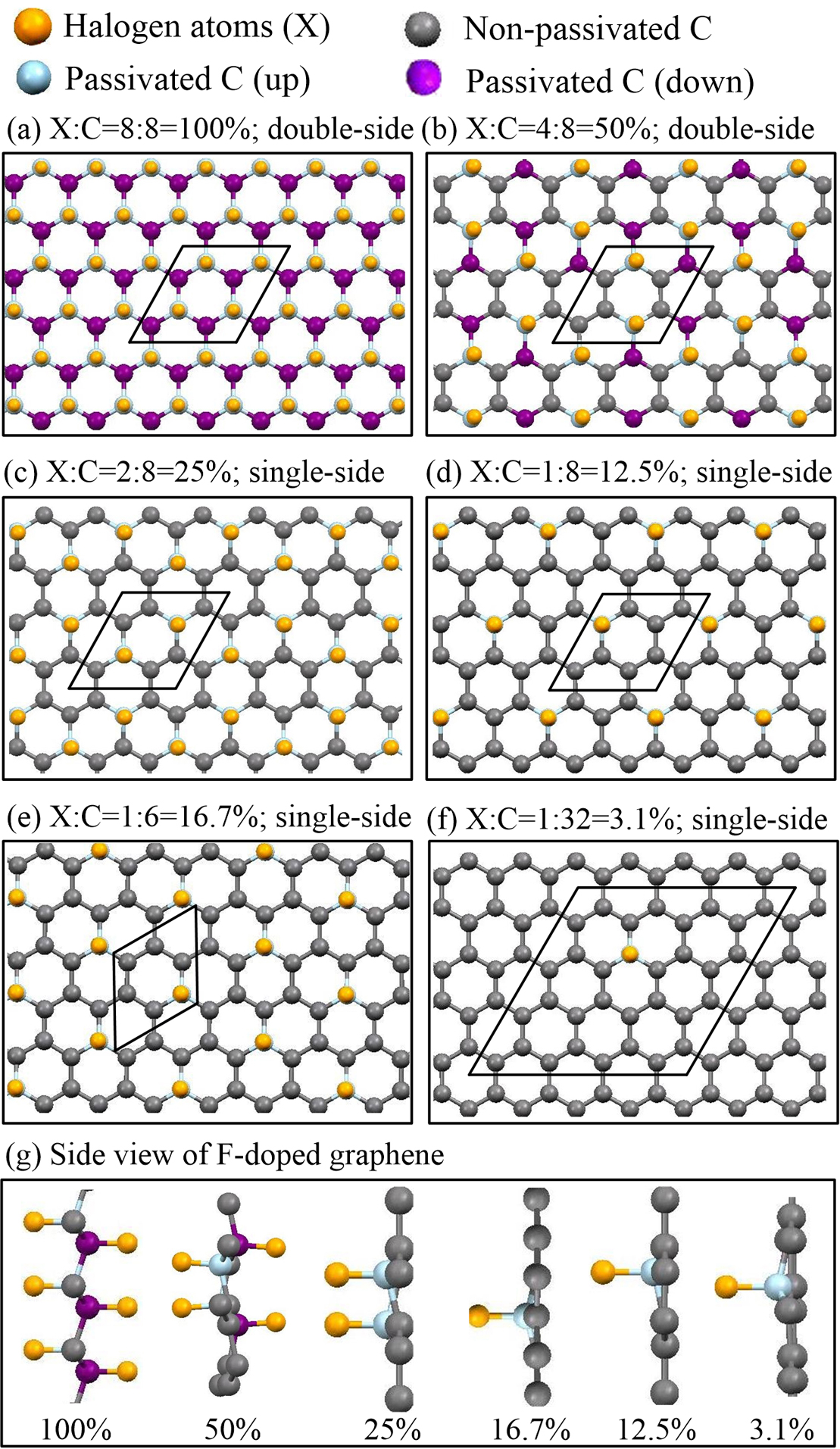}
\centering\caption{}
\end{figure}

\newpage
\newpage
\begin{figure}[hp]
\graphicspath{{figure}}
\centering
\includegraphics[scale=1.3]{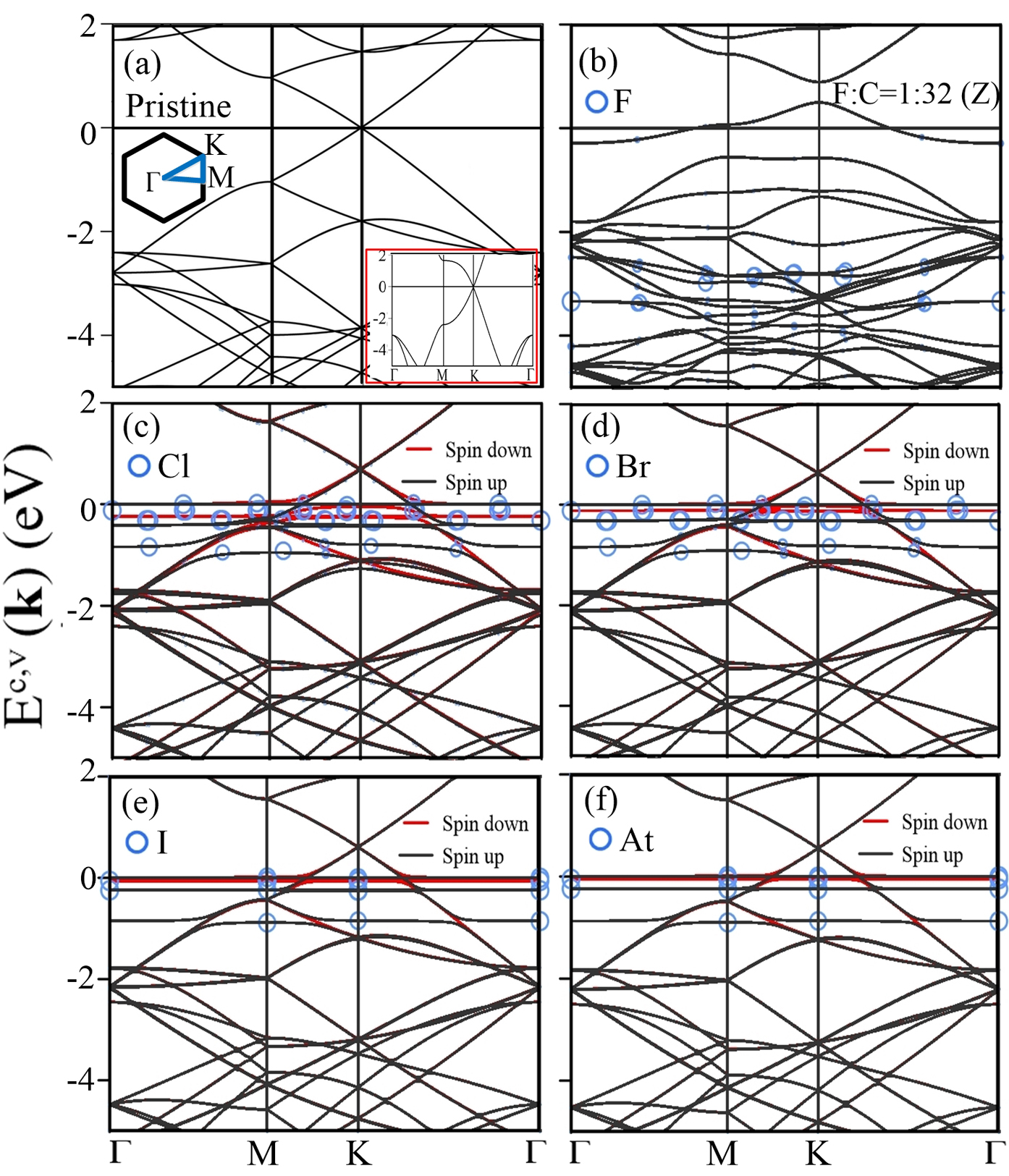}
\centering\caption{}
\end{figure}

\newpage
\newpage
\begin{figure}[hp]
\graphicspath{{figure}}
\centering
\includegraphics[scale=2.5]{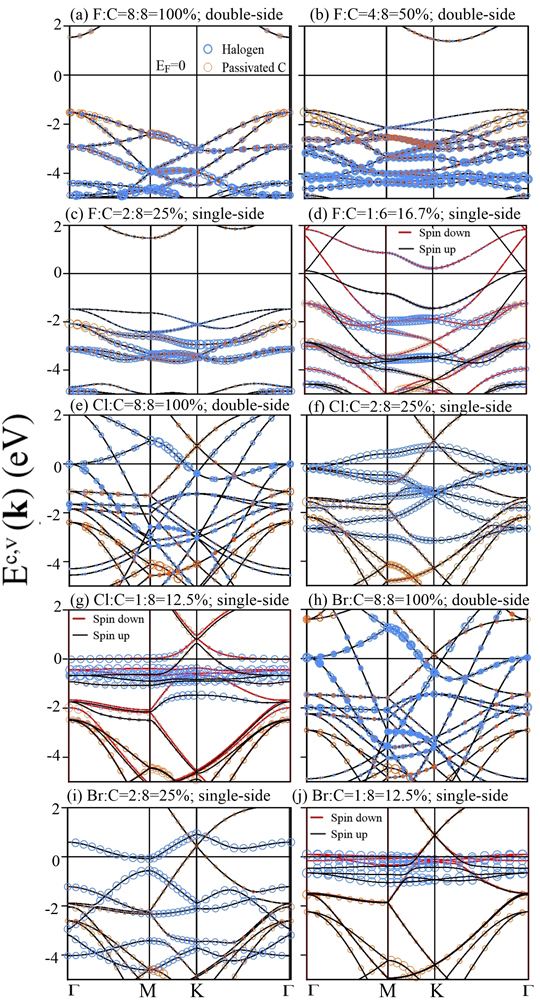}
\centering\caption{}
\end{figure}

\newpage
\newpage
\begin{figure}[hp]
\graphicspath{{figure}}
\centering
\includegraphics[scale=1]{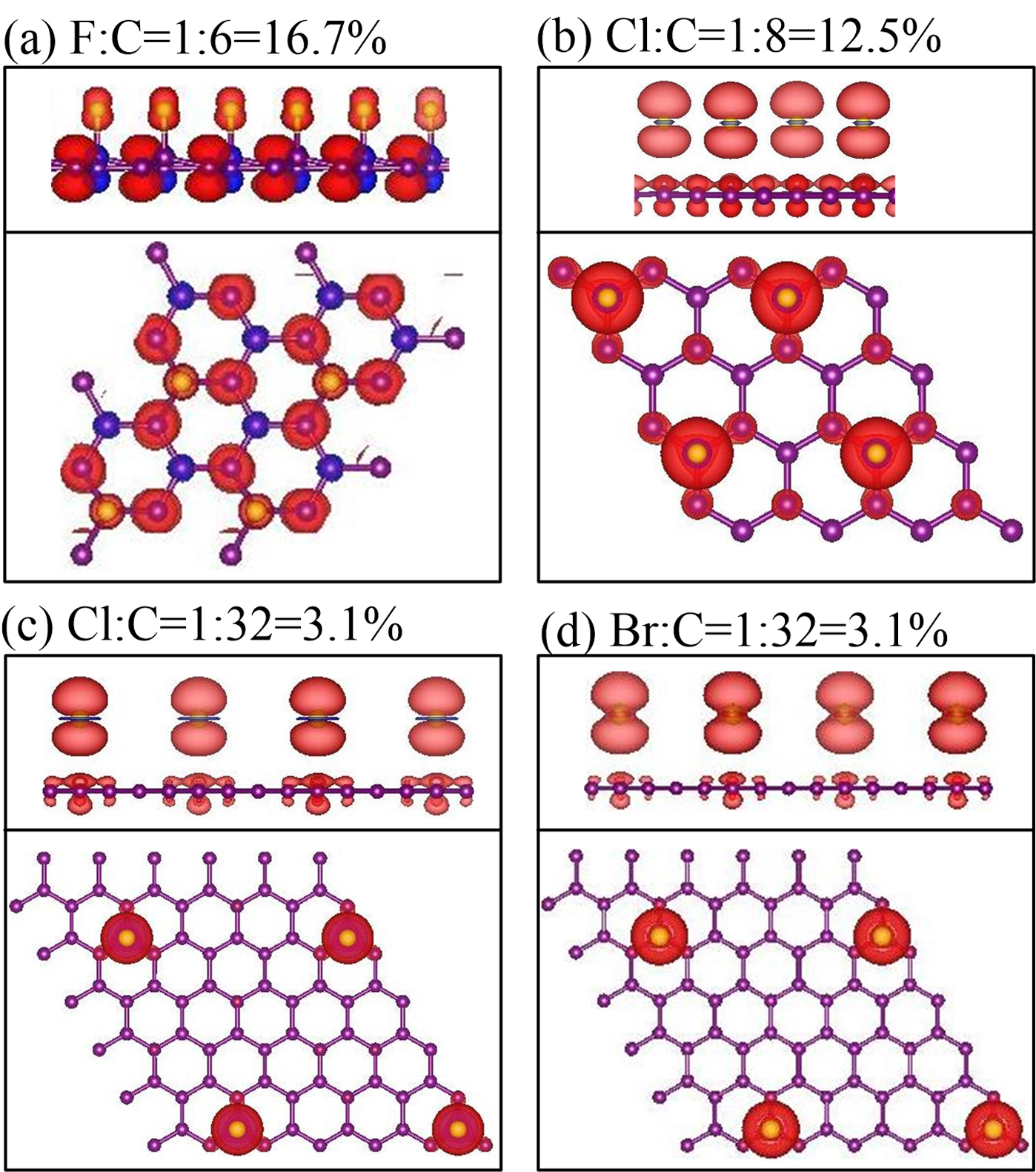}
\centering\caption{}
\end{figure}

\newpage
\newpage
\begin{figure}[hp]
\graphicspath{{figure}}
\centering
\includegraphics[scale=1.2]{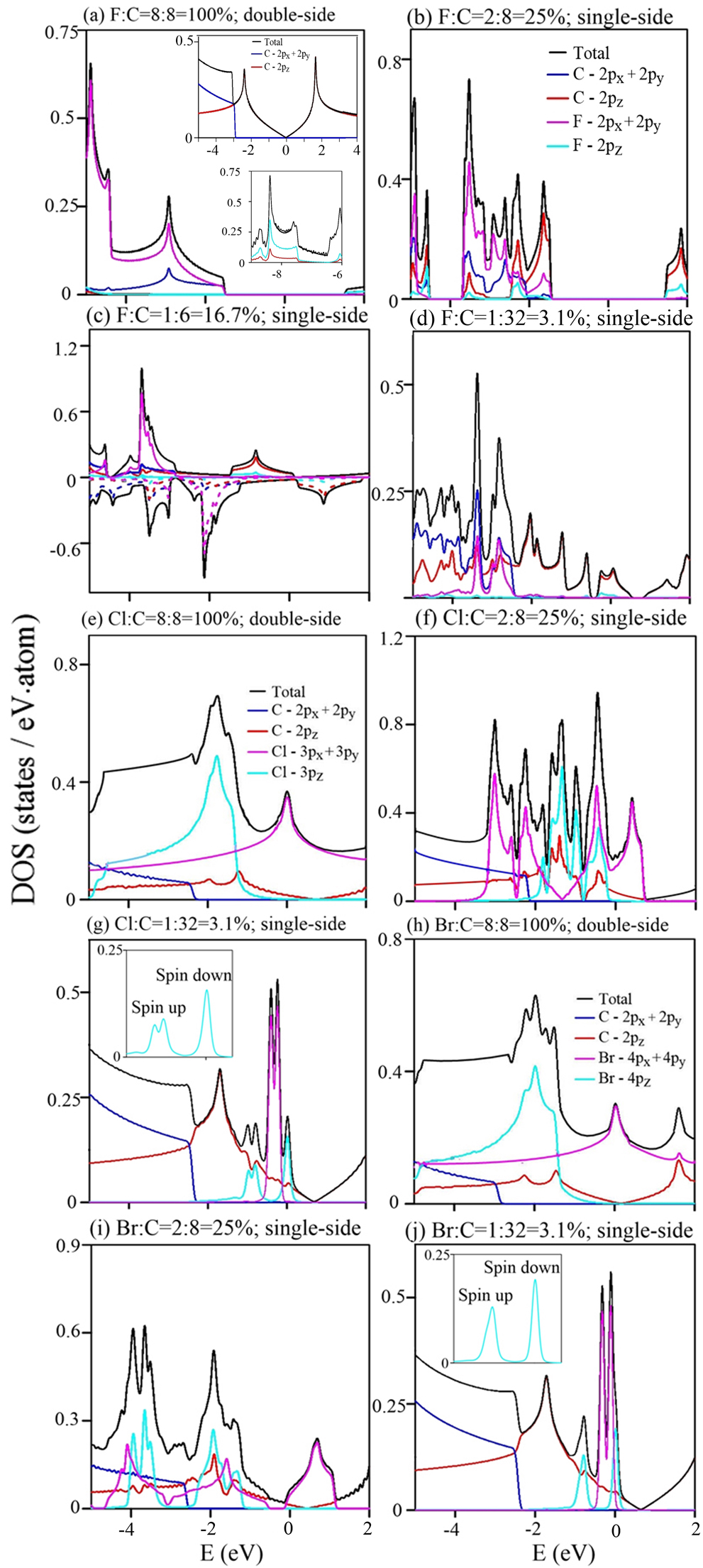}
\centering\caption{}
\end{figure}

\newpage
\newpage
\begin{figure}[hp]
\graphicspath{{figure}}
\centering
\includegraphics[scale=1.2]{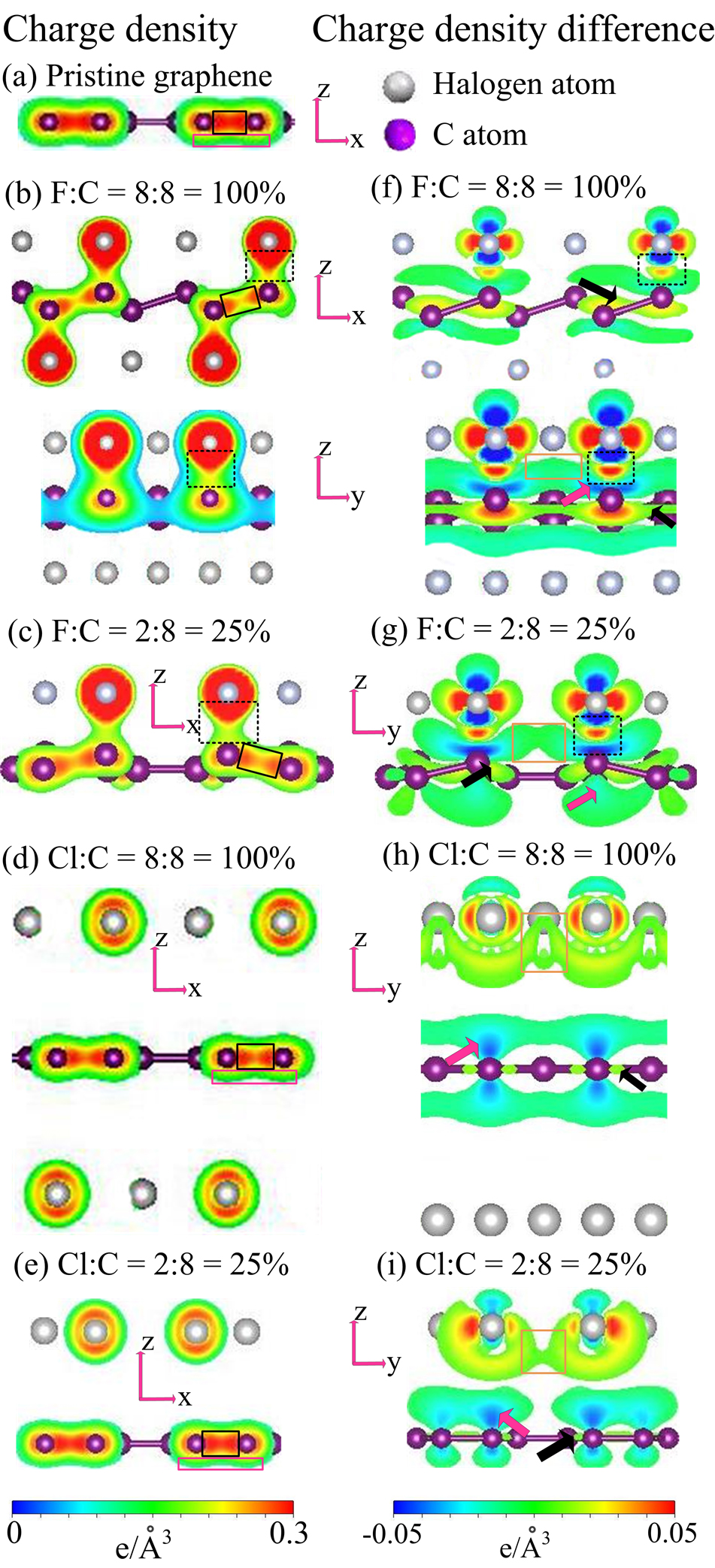}
\centering\caption{}
\end{figure}

\end{document}